\documentclass[twocolumn,showpacs,preprintnumbers,amsmath,amssymb,superscriptaddress]{revtex4}

\usepackage{graphicx}
\begin{document}
\bibliographystyle{prsty}
\title{Manifestation of Correlation Effects in the photoemission spectra of Ca$_{1-x}$Sr$_x$RuO$_3$}

\author{M.~Takizawa}
\affiliation{Department of Physics and Department of Complexity 
Science and Engineering, University of Tokyo, 
5-1-5 Kashiwanoha, Kashiwashi, Chiba, 277-8561, Japan}
\author{D.~Toyota}
\affiliation{Department of Applied Chemistry, University of Tokyo, 
Bunkyo-ku, Tokyo 113-8656, Japan}
\author{H.~Wadati}
\affiliation{Department of Physics and Department of Complexity 
Science and Engineering, University of Tokyo, 
5-1-5 Kashiwanoha, Kashiwashi, Chiba, 277-8561, Japan}
\author{A.~Chikamatsu}
\affiliation{Department of Applied Chemistry, University of Tokyo, 
Bunkyo-ku, Tokyo 113-8656, Japan}
\author{H.~Kumigashira}
\affiliation{Department of Applied Chemistry, University of Tokyo, 
Bunkyo-ku, Tokyo 113-8656, Japan}
\author{A.~Fujimori}
\affiliation{Department of Physics and Department of Complexity 
Science and Engineering, University of Tokyo, 
5-1-5 Kashiwanoha, Kashiwashi, Chiba, 277-8561, Japan}
\author{M.~Oshima}
\affiliation{Department of Applied Chemistry, University of Tokyo, 
Bunkyo-ku, Tokyo 113-8656, Japan}
\author{Z.~Fang}
\affiliation{Institute of Physics, Chinese Academy of Science, Beijing 100080, China}
\author{M.~Lippmaa}
\affiliation{Institute for Solid State Physics, University of Tokyo, 5-1-5 Kashiwanoha, Kashiwashi, Chiba, 277-8581, Japan}
\author{M.~Kawasaki}
\affiliation{Institute for Materials Research, Tohoku University, 
  2-1-1 Katahira, Aoba-ku, Sendai, Miyagi, 980-8577, Japan}
\author{H.~Koinuma}
\affiliation{Materials and Structures Laboratory, Tokyo Institute of Technology, 4259 Nagatsuta, Midori-ku, Yokohama, Kanagawa, 226-8503, Japan}
\date{\today}

\begin{abstract}
We have measured soft x-ray photoemission and O 1{\it s} x-ray absorption spectra of Ca$_{1-x}$Sr$_x$RuO$_3$ thin films prepared {\it in situ}. 
The coherent and incoherent parts have been identified in the bulk component of the photoemission spectra, and spectral weight transfer from the coherent to the incoherent part has been observed with decreasing $x$, namely, with increasing orthorhombic distortion. 
We propose that, while the Ru $4d$ one-electron bandwidth does not change with $x$, the distortion and hence the splitting of the $t_{2\text{g}}$ band effectively increases electron correlation strength. 
Although strong mass enhancement is found in the electronic specific heat data, the coherent part remains wide, suggesting enhanced band narrowing only in the vicinity of {\it E$_{F}$}. 
\end{abstract}

\pacs{71.28.+d, 71.30.+h, 73.61.-r, 79.60.Dp}

\maketitle

Metal-insulator transition has been extensively studied because of its fundamental importance in condensed matter physics as well as of its close relationship with remarkable phenomena such as the high-temperature superconductivity in cuprates and the colossal magnetoresistance in manganites \cite{MIT}. 
Broadly speaking, metal-insulator transition occurs in two ways. 
One is bandwidth control, and the other is filling control. 
In bandwidth control, bandwidth and hence electron correlation strength is changed through the modification of, e.g., the lattice parameters. 
Recent developments of dynamical mean-field theory (DMFT) have led to a lot of progress in understanding many problems inherent in strongly correlated electron systems, including Mott metal-insulator transition \cite{DMFT}. 
According to DMFT, as $U/W$ increases, where $U$ is the on-site Coulomb energy and $W$ is the one-electron bandwidth, spectral weight is transferred from the coherent part (the quasiparticle band near {\it E$_{F}$}) to the incoherent part (the remnant of the Hubbard bands $1 - 2$ eV above and below {\it E$_{F}$}) \cite{Zhang}. 
Metal-to-insulator transition thus occurs as the spectral weight of the coherent part vanishes. 

In perovskite-type {\it AB}O$_{3}$ compounds, bandwidth control is realized through the modification of the radius of the {\it A} site ion $r_{A}$ and hence the modification of the {\it B}-O-{\it B} bond angle. 
As $r_{A}$ decreases, the bond angle decreases from $180^{\circ}$ and the cubic lattice transforms to the orthorhombic (GdFeO$_{3}$-type) structure. 
The orthorhombic distortion reduces $W$, because the effective transfer integrals between the neighboring $B$ sites is governed by the super-transfer process via the O $2p$ state. 
{\it R}NiO$_{3}$ ({\it R} $=$ rare earth) is a typical example of bandwidth control systems and shows a metal-insulator transition as a function of the ionic radius of {\it R} \cite{RNiO}. 
Ca$_{1-x}$Sr$_{x}$VO$_{3}$ is another typical bandwidth control system but remains metallic for the entire $x$ range. 
Inoue {\it et al.} \cite{CSVO-Inoue} have reported that in the photoemission (PES) spectra of Ca$_{1-x}$Sr$_{x}$VO$_{3}$, as one decreases $x$, that is, as one decreases $W$, spectral weight is transferred from the coherent part to the incoherent part centered at $\sim$ 1.5 eV below {\it E$_{F}$}. 
Recently, however, Sekiyama {\it et al.} \cite{CSVO-Sekiyama} have reported that spectral weight transfer is not observed in the PES spectra of Ca$_{1-x}$Sr$_{x}$VO$_{3}$ measured using high photon energies, i.e., in so-called "bulk-sensitive" photoemission spectra. 
Very recently, a more bulk-sensitive measurement of the same system using a UV laser as a light source indicated that spectral weight transfer indeed occurs \cite{CSVO-Eguchi}. 
Unfortunately, accessible energy range in the UV laser PES experiment is limited to $\sim$ 1 eV below {\it E$_{F}$} and therefore the entire incoherent part cannot be observed. 
The problem therefore remains highly controversial and further studies are strongly required. 
In the present work we have observed clear spectral weight transfer in the bandwidth-control system Ca$_{1-x}$Sr$_x$RuO$_3$ (CSRO) after decomposition of the spectra into surface and bulk components. 

Despite their more extended nature of the $4d$ orbitals than the transition-metal $3d$ orbitals, ruthenates are found to show various phenomena associated with electron correlation such as unconventional superconductivity in Sr$_{2}$RuO$_{4}$ \cite{maeno214} and metal-insulator transition in Ca$_{2-x}$Sr$_{x}$RuO$_{4}$ \cite{nakatsuji214}. 
CSRO studied in the present work is metallic in the entire $x$ range. 
SrRuO$_{3}$ is metallic and shows ferromagnetism below $T_{\text{C}} \simeq 160$ K \cite{SRO-Tc}. 
In going from SrRuO$_{3}$ to CaRuO$_{3}$, the Ru-O-Ru bond angle is reduced from $\sim$ $165^{\circ}$ to $\sim$ $150^{\circ}$ \cite{Ru-O-Ru} and $T_{\text{C}}$ decreases to zero at $x \sim 0.4$ \cite{kanbayashi}. 
Optical studies of CSRO have shown characteristic behavior of a Mott-Hubbard system predicted by DMFT \cite{SCRO-Opt}. 
The anomalous Hall effect of SrRuO$_{3}$ has been explained as due to the presence of magnetic monopoles in the momentum space \cite{monopole}. 

Previous photoemission studies of polycrystalline SrRuO$_{3}$ samples have revealed that the effect of electron correlation is substantial within the Ru $4d$ $t_{2\text{g}}$ band \cite{okamoto, fujioka}. 
Recent Ru $3d$ core-level x-ray photoemission studies of various ruthenates \cite{Ru3d} have also revealed the importance of electron correlation effects in the ruthenates. 
Recently, the growth of high-quality perovskite oxide single-crystal thin films has become possible by pulsed laser deposition (PLD) \cite{Izumi,Choi}, and a setup has been developed for their {\it in-situ} photoemission measurements \cite{Horiba,Wadati}.
In this work, we have measured soft x-ray PES and absorption spectra of CSRO thin film. 
Ultra-violet PES experiment of {\it ex-situ} prepared CSRO thin films \cite{sjoh-CSROfilm} has also been reported recently.

The PES and x-ray absorption spectroscopy (XAS) measurements were performed at BL-2C of Photon Factory (PF), High Energy Accelerators Research Organization (KEK), using a combined laser molecular beam epitaxy (MBE)-photoemission spectrometer system. Details of the experimental setup are described in Ref \cite{Horiba}. Epitaxial films of CSRO were grown on single-crystal substrates of Nb-doped SrTiO$_3$ by the PLD method. 
The substrates were annealed at 1050${}^{\circ}$C under an oxygen pressure of $\sim 1\times 10^{-6}$ Torr to obtain an atomically flat TiO$_2$-terminated surface \cite{kawasaki}. 
CSRO thin films were deposited on the substrates at 900${}^{\circ}$C at an oxygen pressure of $\sim 0.1$ Torr. 
The surface morphology of the measured films was checked by {\it ex-situ} atomic force microscopy (AFM), showing atomically flat step-and-terrace structures. 
Coherent growth on the substrate was confirmed by x-ray diffraction. 
Due to the coherent growth, the out-of-plane lattice constant decreased with Ca doping while the in-plane lattice constant did not vary. 
Therefore the Ru-O-Ru bond angle is thought to be reduced in going from SrRuO$_{3}$ to CaRuO$_{3}$ in the CSRO thin films, too. 
The fabricated CSRO thin films showed metallic resistiviy and the values of their $T_{\text{C}}$ were almost the same as those of the bulk samples \cite{kanbayashi}. 
All the photoemission measurements were performed under an ultrahigh vacuum of $\sim 10^{-10}$ Torr at room temperature using a Scienta SES-100 electron-energy analyzer. The total energy resolution was about 150 meV at the photon energy of 400 eV, about 200 meV at that of 600 eV and about 500 meV at that of 900 eV. The Fermi level ($E_F$) position was determined by measuring gold spectra. The XAS spectra were measured in the total-electron-yield mode.

\begin{figure}
\begin{center}
\includegraphics[width=7cm]{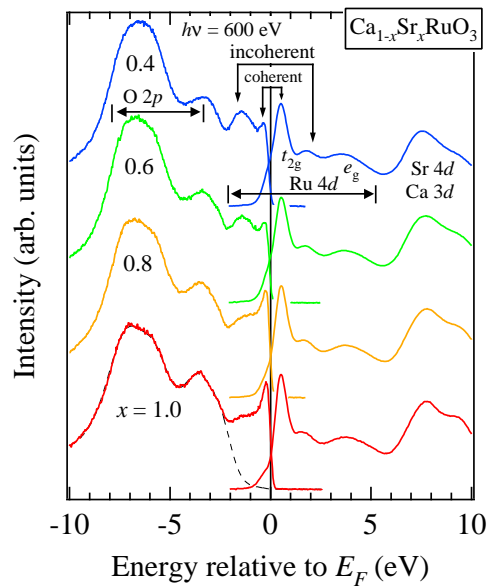}
\caption{(Color online) A combined plot of the valence-band spectra and the O 1$s$ XAS spectra of Ca$_{1-x}$Sr$_x$RuO$_3$. The dashed curve shows how the O $2p$ band has been subtracted to obtain the Ru $4d$ band. }
\label{VB}
\end{center}
\end{figure}

The Ru 3$d$ core level (not shown) reveals two sets of peaks, well-screened peaks and poorly-screened peaks and the ratio of the poorly-screened to well-screened peaks increased with increasing Ca content, in agreement with Cox {\it et al.} \cite{Ru-Cox}. 
Figure~\ref{VB} shows a combined plot of the valence-band spectra and the O 1$s$ XAS spectra, representing the electronic state below and above {\it E$_{F}$}. 
The band between $\sim -2$ eV to $\sim 5$ eV measured relative to {\it E$_{F}$} is mainly composed of Ru $4d$ character and that between $\sim -10$ eV to $\sim -2$ eV of O $2p$ character \cite{okamoto,fujioka}. 
For PES, we assign the emission within $\sim 1$ eV of {\it E$_{F}$} with a sharp Fermi edge and the broad band centered at $\sim -1.2$ eV peak, respectively, to the coherent and incoherent parts of the spectral function. 
Similarly for O $1s$ XAS, two peaks are seen within $\sim 2$ eV above the threshold, which we attribute to the coherent and incoherent parts of the unoccupied $t_{2\text{g}}$ band. 
Above the $t_{2\text{g}}$ band, one can see the empty $e_{\text{g}}$ band centered at $\sim 4$ eV. 
Both in PES and XAS, one can see that spectral weight transfer occurs from the coherent part to the incoherent part with Ca doping. 
In XAS, spectral weight transfer is relatively small partly because the core-hole potential distorts the single-particle excitation spectrum and also because surface effects are much weaker in XAS than in PES, as described below. 

\begin{figure}
\begin{center}
\includegraphics[width=6.8cm]{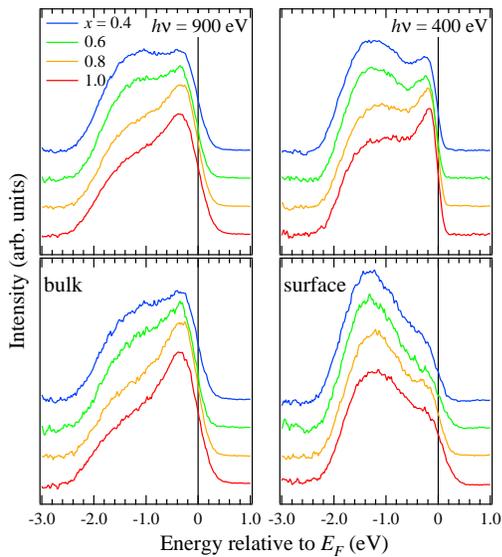}
\caption{(Color online) Ru $4d$ band of Ca$_{1-x}$Sr$_x$RuO$_3$. Top: Data taken at $h\nu = 400$ and 900 eV. Bottom: The bulk and surface components deduced from different energies. }
\label{d-bulksurf}
\end{center}
\end{figure}

Although our PES spectra are relatively bulk-sensitive due to the use of soft x-rays, surface effects are not negligible. 
On the surface, correlation effect is generally enhanced due to the reduced atomic coordination and hence the reduced bandwidth.
Therefore, to study the intrinsic bulk electronic structure, we need to subtract the surface components from the measured spectra following the procedure of Ref.~\cite{LCVO-maiti}. 
The measured photoemission spectra are assumed to be expressed as $I(E) = \exp (-s/\lambda)I_{\text{bulk}}(E) + \Bigl[1 - \exp (-s/\lambda)\Bigr]I_{\text{surface}}(E)$, where $s$ is the thickness of the surface layer, $\lambda$ is the photoelectron mean free path, and $I_{\text{bulk}}$ and $I_{\text{surface}}$ denote the spectra of the bulk and surface regions, respectively. 
Because $\lambda$ is a function of photon energy, one can obtain the bulk and surface components separately from PES spectra measured at two photon energies. 
Here, since we measured atomically flat surfaces, the use of the above formula becomes more reliable. 
We measured PES spectra at different 400 eV and 900 eV as shown in Fig.~\ref{d-bulksurf}. 
As we are interested in the Ru $4d$ band, the tail of the O $2p$ band has been subtracted as shown in Fig.~\ref{VB} and the resulting Ru $4d$ spectra have been normalized to the integrated intensity from $E = 0.5$ to $-3.0$ eV, as shown in Fig.~\ref{d-bulksurf}. 
In order to obtain the bulk and surface components of the Ru $4d$ band, we have used the mean free-paths of $\lambda _{900} = 18$ \AA ~and $\lambda _{400} = 6$ \AA ~at $h\nu = 900$ and 400 eV, respectively according to Ref. \cite{Tanuma-mfp} and the surface layer thickness of 4 \AA, the dimension of the unit cell \cite{Ru-O-Ru}. 
As shown in Fig.~\ref{d-bulksurf}, the incoherent part of the bulk component thus obtained is weaker than that of the raw spectra, and manifests itself as a shoulder rather than a separate feature from the main structure near $E_{F}$. 
Nevertheless, there is still spectral weight transfer from the coherent part to the incoherent part with Ca doping (Fig. \ref{d-bulksurf}). 
The incoherent part is strong in the surface components as expected, similar to the spectra taken at low photon energies \cite{sjoh-CSROfilm}. 

\begin{figure}
\begin{center}
\includegraphics[width=5.5cm]{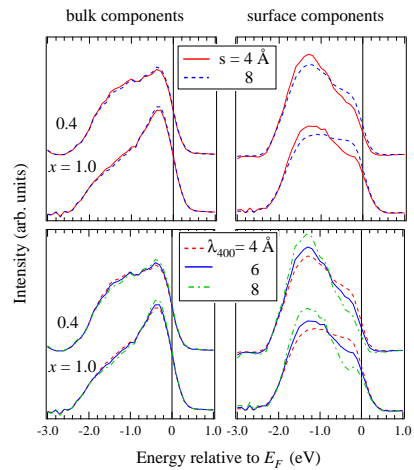}
\caption{(Color online) Bulk (left) and surface (right) components of Ca$_{1-x}$Sr$_{x}$RuO$_{3}$ deduced for a various sets of parameters. $s = 4$ \AA, $\lambda _{400}= 6$ \AA, $\lambda _{900}= 18$ \AA  ~otherwise stated. When $\lambda _{900}$ is varied from 14 to 22 \AA, the bulk and surface components hardly change. }
\label{par}
\end{center}
\end{figure}
In order to see how the result is robust against the uncertainties in the parameters $\lambda$ and $s$, we have varied these parameters as shown in Fig.~\ref{par}. 
The bulk components do not change their line shapes significantly for the range of the parameters $\lambda_{400}$ and $s$. 
The result is even less sensitive to $\lambda_{900}$. 
Therefore, we consider that the obtained bulk component indeed represent the intrinsic bulk electronic structure. As for the surface component, on the other hand, the line shapes are somewhat more sensitive to the parameters. 

\begin{figure}
\begin{center}
\includegraphics[width=6.5cm]{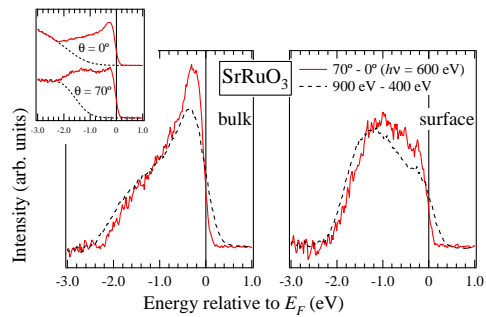}
\caption{(Color online) Bulk and surface components of SrRuO$_{3}$ deduced from the angle dependence of PES spectra. The dashed curves show the surface and bulk components from Fig.~\ref{d-bulksurf}. Inset: raw spectra of SrRuO$_{3}$ taken at $h\nu = 600$ eV. }
\label{angle}
\end{center}
\end{figure}

We have also examined the angle dependence of PES spectra to separate the bulk and surface components. 
From the PES spectra taken at emission angles $0^{\circ}$ and $70^{\circ}$, we have obtained each component as shown in Fig.~\ref{angle}. 
While the surface component is a little different from that obtained from the photon energy dependence (dashed curve), the bulk component is very similar to it. 
(The broader Fermi cut-off in the dashed curve is due to the lower resolution of the 900 eV spectra.) 
This gives additional evidence that the obtained bulk components represent the intrinsic bulk electronic structure. 

\begin{figure}
\begin{center}
\includegraphics[width=5cm]{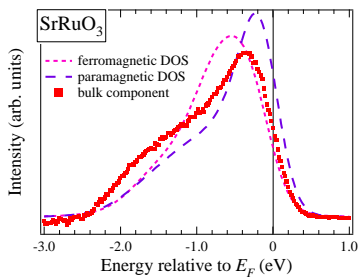}
\caption{(Color online) Comparison of the bulk component and the band-structure calculation for ferromagnetic and paramagnetic SrRuO$_3$. The calculated DOS is broadened with a Gaussian and a Lorentzian function. }
\label{FMPM}
\end{center}
\end{figure}

Since all the measurements were made at room temperature, all the Ca$_{1-x}$Sr$_{x}$RuO$_{3}$ samples were in the paramagnetic state \cite{CaoTc}. 
However, because the $T_{\text{C}}$ decreases from 160 K for $x = 1$ to $\sim$0 K for $x = 0.4$ \cite{kanbayashi}, the spectral change could be due to changes in the ferromagnetic fluctuations, which become stronger with Sr content, rather than changes in the correlation strength. 
In order to see whether this is the case or not, we compare in Fig.~\ref{FMPM} the experiment with the calculated density of states (DOS) of SrRuO$_{3}$ in the ferromagnetic and paramagnetic states \cite{Fang-calc}. 
The calculated DOS for the paramagnetic state clearly shows more pronounced peak at $E_{F}$ contrary to the $x$-dependent change in the spectra. 
This would exclude the magnetic fluctuations as the origin of the spectral weight transfer. 

Since CaRuO$_{3}$ is more distorted than SrRuO$_{3}$, one would naturally expect the narrowing of the Ru $4d$ band in going from SrRuO$_{3}$ to CaRuO$_{3}$. 
However, the band-structure calculation has shown \cite{CSRO-bandcal-mazin} that the band width of Ru $4d$ does not change between SrRuO$_3$ and CaRuO$_3$. 
Although the orthorhombic distortion does not reduce the $t_{2\text{g}}$ bandwidth, it lifts the $t_{2\text{g}}$ orbital degeneracy. 
Such a lift of the $t_{2\text{g}}$-orbital degeneracy has been shown to effectively increase the electron correlation strength in the case of Ti perovskites \cite{pavarini}. 
The spectral weight transfer from the coherent part to the incoherent part with Ca doping would therefore be due to the increased correlation strength due to the lifting of the orbital degeneracy. 

Compared with the band-structure calculation, the bandwidth of experimental "bulk" data is similar to that of calculation. 
This may mean that $m^*/m_b$ is $\sim 1$, where $m^*$ is the effective mass of the quasiparticle and $m_b$ is the bare band mass, if the band narrowing occurs uniformly in the entire Ru $4d$ band. 
However, specific heat measurement indicated that the value of $m^*/m_b$ is about 10 for SrRuO$_3$ and 17 for CaRuO$_3$ \cite{CaoTc}. This implies that $m^*$ is enhanced in the vicinity of $E_F$, which in turn means that the genuine coherent part exists only in the vicinity of $E_F$. This is analogous to the case of U compounds (UAl$_2$, UPt$_3$) \cite{UAl} where the $5f$ band width is similar or larger than the band-structure calculation but the mass enhancement at $E_F$ can be as large as $m^*/m_b \sim 100$. These features are consistent with the bad metallic behavior of SrRuO$_3$ \cite{badmetal-SRO}.

In conclusion, with bandwidth control, spectral weight transfer is observed between the coherent and incoherent parts. 
As the orthorhombic distortion lifts the $t_{2\text{g}}$ orbital degeneracy, the effective correlation strength is increased. 
The similarity to U compounds in the spectrum of SrRuO$_3$ suggests that the spectral changes from $4f$ (Ce) to $5f$ (U) electron systems may correspond to that from $3d$ to $4d$ electron systems, namely well separated coherent and incoherent features to unresolved ones. 

The authors would like to thank A.~Liebsch and T.~Yoshida for enlightening discussions and K.~Ono and A.~Yagishita for their support in the experiment at PF. 
This work was supported by a Grant-in-Aid for Scientific Research (A16204024) from the Japan Society for the Promotion of Science. 
The work was done under the approval of Photon Factory Program Advisory Committee (Proposal No. 2003G149) and Project No. 2003G149 at the Institute of
Material Structure Science, KEK. 
Z.~Fang acknowledges the supports from NSF of China.

\end{document}